\documentclass[twocolumn]{aastex631}

\shorttitle{Fermi Radio Multi-Survey}
\shortauthors{Bruzewski et al.}

\def\gammaray{$\gamma$-ray }
\def\gammarays{$\gamma$-rays }

\begin{document}

\received{2022-10-06}
\accepted{2022-12-09}
\submitjournal{ApJ}

\title{A Combined Radio Multi-Survey Catalog of Fermi Unassociated Sources}

\author[0000-0001-7887-1912]{S. Bruzewski}
\affiliation{Department of Physics and Astronomy, University of New Mexico, Albuquerque, NM 87131, USA}

\author[0000-0001-6672-128X]{F.K. Schinzel}
\altaffiliation{An Adjunct Professor at the University of New Mexico.}
\affiliation{National Radio Astronomy Observatory, P.O. Box O, Socorro, NM 87801, USA}

\author[0000-0001-6495-7731]{G.B. Taylor}
\affiliation{Department of Physics and Astronomy, University of New Mexico, Albuquerque, NM 87131, USA}


\correspondingauthor{S. Bruzewski}
\email{bruzewskis@unm.edu}

\begin{abstract}
    Approximately one-third of existing \gammaray sources identified by the \textit{Fermi Gamma-Ray Space Telescope} are considered to be unassociated, with no known counterpart at other frequencies/wavelengths. These sources have been the subject of intense scrutiny and observational effort during the observatory's mission lifetime, and here we present a method of leveraging existing radio catalogs to examine these sources without the need for specific dedicated observations, which can be costly and complex. Via the inclusion of many sensitive low-frequency catalogs we specifically target steep spectrum sources such as pulsars. This work has found steep-spectrum radio sources contained inside 591 \textit{Fermi} unassociated fields, with at least 21 of them being notable for having pulsar-like \gammaray properties as well. We also identify a number of other fields of interest based on various radio and \gammaray selections.
\end{abstract}

\keywords{High energy astrophysics (739) -- Surveys (1671) -- Radio source catalogs (1356) -- Spectral index (1553) -- Radio astronomy (1338) -- Active galaxies (17) -- Gamma-ray astronomy (628) -- Radio
continuum emission (1340)}

\section{Introduction} \label{sec: intro}

During its mission lifetime, the \textit{Fermi Gamma-Ray Space Telescope} has provided substantial insights into the high energy regime of the sky. Of particular note are the resolved \gammaray sources described in each subsequent data release from the Large Area Telescope \citep{2009ApJ...697.1071A}. The most recent release \citep[4FGL-DR3; see][]{2022ApJS..260...53A} features some 6659 sources, many falling into categories one might expect, such as pulsars or active galactic nuclei (AGN). Beyond these sources though, the catalog has presented a persistent mystery in the form of its unassociated sources: \gammaray sources which are detected by \textit{Fermi}, but cannot be definitively associated with or identified as a source in any other electromagnetic regime. 

Currently the number of such sources stands at 2157, making up 32\% of the entire catalog. As \gammaray properties alone are generally insufficient to determine the physical nature of a source, this means that nearly 1/3 of all presently known astrophysical \gammaray sources have an unknown origin. One should also note that the existence of these sources cannot be completely explained away by invoking signal-to-noise arguments. While present unassociated sources do generally have somewhat lower fluxes, there are a non-insignificant number of bright unassociated sources which have persisted across catalog updates over the years. As an example, the source 4FGL J1801.6-2326 was originally published in the very first FGL catalog (0FGL), and has remained unassociated even though it is a "bright" source by any criteria which one might imagine. This source is, at most, loosely associated with the supernova remnant W28, although there are other \gammaray sources which are more definitively linked to that object \citep{2010ApJ...718..348A}.

The existence of such persistent sources implies that some fraction of the unassociated sources will continue to remain inscrutable against our traditional methods of association and identification, and thus necessitates the use of new methods which might be able to provide insight into these unassociated fields. The radio regime has long served as a primary testing-ground for such methods, as most sources of \gammarays (e.g. AGN and pulsars) would typically be expected to appear in the radio (with the exception of a few exotic systems), and the smaller sky-density of sources compared to optical or near-optical makes the problem somewhat more tractable. The techniques used in radio observations thus far can largely be divided into two categories: association via statistical likelihood arguments based on source properties (common for AGN), and identification via correlated variability (common for pulsars). Both of these methods have found remarkable success, with a majority of sources in both 4LAC and 2PC (the \textit{Fermi} catalogs of AGN and pulsars, see \citet{2020ApJ...892..105A} and \citet{2013ApJS..208...17A} respectively) having been originally identified/associated via radio observations.

For previous searches (such as those described in \citet{2015ApJS..217....4S} and \citet{2013MNRAS.432.1294P}), each new list of unassociated sources was targeted using the NSF's Karl G. Jansky Very Large Array (VLA) at 5 and 7 GHz, looking for bright, compact sources inside the positional uncertainty ellipse. These candidate sources could then be targeted using Very Long Baseline Interferometry (VLBI), typically via the Very Long Baseline Array (VLBA). If these follow-up observations detected the source, one could then apply a statistical argument: what is the likelihood this source is the \gammaray emitter, as opposed to a chance background source, knowing that we have near completeness on said compact radio sources at these frequencies thanks to the calibrator surveys such as \citet{2021AJ....161...14P}\footnote{Radio Fundamental Catalog: \url{http://astrogeo.org/rfc/}}? If this quantified likelihood surpasses some threshold, the source may be considered a candidate for association, and passed along to other groups for further analysis. 

A new approach began to be implemented with the release of the 8-year catalog \citep[4FGL,][]{2020ApJS..247...33A}. By cataloging sources in the early data products of the recently completed first epoch of the Very Large Array Sky Survey \citep[VLASS,][]{2020PASP..132c5001L} we were able to link any new 5 and 7 GHz data we acquired to a new data point at 3 GHz, effectively doubling our spectral coverage, and allowing us access to certain \textit{Fermi} unassociated sources which we previously would not have targeted due to observational constraints. A full description of these efforts can be found in \citet{2021ApJ...914...42B}, but to summarize: sources at each frequency are linked using an uncertainty weighted distance metric, such that a graph is generated for each unassociated \textit{Fermi} source. We can then extract all connected component sub-graphs, considering them multi-frequency sources, and fit for their spectral features. 

This work represents the logical extension of that effort to include a larger number of catalogs, thus spanning both an extended range of frequency and a larger portion of the entire sky. In particular, this approach incorporates an increased number of low-frequency radio catalogs, which is ideal if one is searching for sources with steeply negative spectral indices (positive convention, $S_\nu \propto \nu^{+\alpha}$). We highlight these sources because it is well-known that continuum emission from pulsars typically occurs in the range $\alpha<-1.4$ \citep{2013MNRAS.431.1352B}, quite outside the standard range of spectral indices for extragalactic objects \citet{1971BAAS....3R.447C}. 

This continuum based analysis, inspired by similar but more limited efforts such as \citet{2018MNRAS.475..942F} and \citet{2014ApJS..213....3M}, has so far served as the basis of many different observational techniques towards \textit{Fermi} unassociated sources, and has provided a significant number of newly identified \gammaray pulsars and newly associated blazars \citep[see][]{2016A&A...588A.141G}. Furthermore, it should be noted that compared to pulsation searches, this method of pulsar identification is much less sensitive to the effects of pulse scattering by the interstellar medium. A population of scattered \gammaray pulsars is of particular interest given current questions about the nature of the Galactic Center Excess \citep{2011PhLB..697..412H}. Initially this anomalous excess of \gammarays was ascribed to the theoretical annihilation of dark matter particles in the galactic center region \citep{2014PhRvD..89k5022B}, but more recent analysis has produced some contention. While dark matter remains a popular explanation for its origin \citep[see][]{2017ApJ...840...43A,2022MNRAS.511L..55G}, it seems this excess could also be explained by the existence of an unresolved population of \gammaray pulsars in that region. As such, confirmation of this existence (or non-existence) of these objects would be likely to lead to significant progress in our understanding of the Galactic Center Excess. 

The paper is organized as follows: in Section \ref{sec: methods} we describe the process of assembling the various catalogs into a combined system. In Section \ref{sec: analysis} we provide analysis of the connected component sources which we have identified. Finally in Section \ref{sec: discuss} we discuss broader implications of this work. 

\section{Methodology} \label{sec: methods}

The novel approach this paper describes involves the introduction of several new catalogs to an existing source-finding framework established in \citet{2021ApJ...914...42B}. In particular, we sought out any sky-survey which covered a substantial portion of the sky, and provided some new information in frequency or sky coverage space. The end goal was to produce reasonable sky coverage at as many frequencies as possible. 

{\catcode`\&=11\gdef\tabrefa{\citet{2022A&A...659A...1S}}}
{\catcode`\&=11\gdef\tabrefb{\citet{2017A&A...598A..78I}}}
{\catcode`\&=11\gdef\tabrefc{\citet{1997A&AS..124..259R}}}
{\catcode`\&=11\gdef\tabrefd{\citet{2002A&A...394...59D}}}

\vspace*{-2\baselineskip}
\begin{deluxetable*}{lccll}
    \tablecaption{Selected Catalogs}
    \tablehead{\colhead{Catalog} & \colhead{Center Frequency} & \colhead{Number of Sources} & \colhead{Sky Coverage} & \colhead{Reference}} 
    \startdata
    AT20G     & 5, 8, and 20 GHz & 3797, 3795, and 5877 & Southern sky & \citet{2008MNRAS.384..775M} \\
    CGPS      & 1.4 GHz          & 72787                & Northern galactic & \citet{2003AJ....125.3145T} \\
    Dedicated & 5 and 7 GHz      & 12999 and 10247      & Selected Northern sources & \citet{2021ApJ...914...42B} \\
    FIRST     & 1.40 GHz         & 946432               & Northern extra-galactic & \citet{1995ApJ...450..559B} \\
    GLEAM     & 200 MHz          & 329487               & Southern extra-galactic & \citet{2017MNRAS.464.1146H} \\
    LOTSS     & 144 MHz          & 300098               & Selected Northern field & \tabrefa  \\
    NVSS      & 1.40 GHz         & 1773484              & Northern sky & \citet{2006AAS...209.9709K} \\
    PMN       & 4.85 GHz         & 50814                & Southern sky & \citet{1994ApJS...91..111W} \\
    SUMSS     & 843 MHz          & 211047               & Southern extra-galactic & \citet{2003MNRAS.342.1117M} \\
    TGSS      & 1.50 GHz         & 623604               & Northern sky & \tabrefb \\
    VLASS     & 3.00 GHz         & 2232725              & Northern sky &  \citet{2020PASP..132c5001L} \\
    VLITE     & 364 MHz          & 39957                & Selected Northern sources & \citet{2016arXiv160303080C} \\
    VLSSR     & 74.0 MHz         & 92965                & Northern sky & \citet{2014MNRAS.440..327L} \\
    WENSS     & 325 MHz          & 229418               & Northern sky & \tabrefc \\
    WISH      & 352 MHz          & 90357                & Selected Southern field & \tabrefd
    \enddata
    \tablecomments{A brief summary of the various catalogs included in our analysis. For a full description of each catalog, one should see the various provided references.}
\end{deluxetable*} \label{tbl: catalogs}

\begin{figure}
    \centering
    \includegraphics[width=\linewidth]{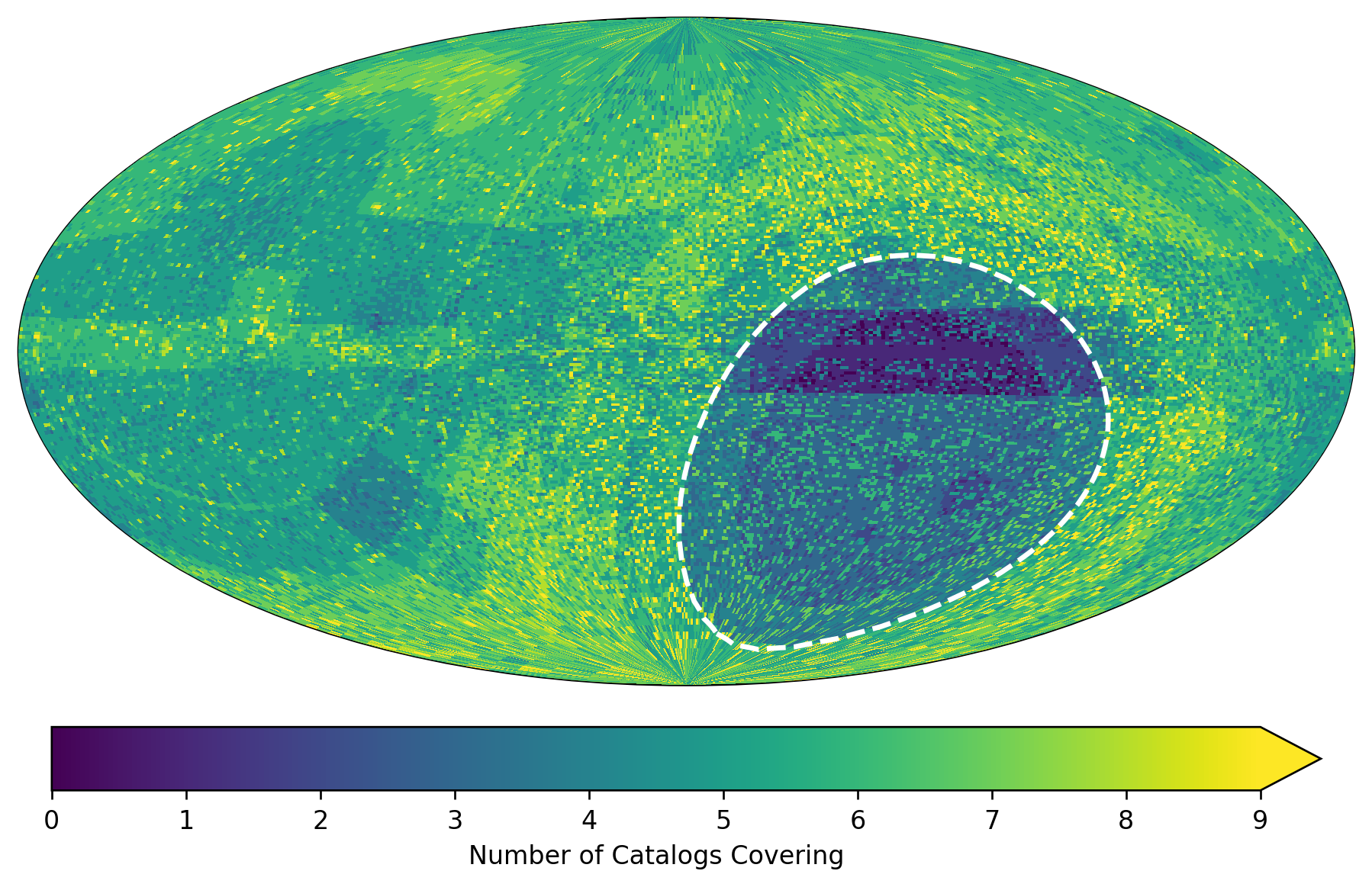}
    \caption{An approximate map of our sky coverage in equatorial coordinates. For each $5^\circ$x $5^\circ$ bin, we count catalogs having at least one source in that region. Note that only a small portion of the Southern Sky is not covered by our catalogs.}
    \label{fig: coverage}
\end{figure}

To this end we identified 14 catalogs which could be added to our existing dedicated observations. A list of these catalogs, along with their general properties, can be seen in Table \ref{tbl: catalogs}. In total we have collected approximately 7 million point sources spanning the frequency range from 74 MHz to 20 GHz. While many of these surveys covered the entire sky available to the instruments involved, some were more specifically targeted, giving us increased coverage in some areas, particularly the galactic plane. It should also be noted that because of the comparative number of instruments in the Southern Hemisphere, our coverage in that region of the sky is somewhat less than the Northern sky. Figure \ref{fig: coverage} shows a summary map of our sky coverage. 

At this stage we perform a similar friends-of-friends graph analysis to identify entries in catalogs which are likely associated with the same physical source. To do this we narrow each catalog to only entries inside of the \textit{Fermi} positional uncertainty ellipse, checking its normalized radius 

\begin{equation}
    \hat{r} = d_S \sqrt{\left(\frac{\cos{\theta_H}}{a_F}\right)^2 + \left(\frac{\sin{\theta_H}}{b_F}\right)^2} \leq 1 \text{,}
\end{equation}

where $d_S$ is the distance from the center of the ellipse to the source, $\theta_H$ is the angle between the ellipse major axis and the source, and $a_F$ and $b_F$ are the semi-major and semi-minor axes of the ellipse, respectively. It is necessary to select for interior sources at this stage because the graphing analysis scales roughly as $\mathcal{O}(n_{sources}^2)$, making a whole-sky analysis of 7 million sources computationally prohibitive. We instead select for nearby sources (a relativity inexpensive calculation) and then perform the analysis on at most a few dozen sources inside the ellipse. 

Once we have our list of interior sources we can begin to network them. For catalogs generated using the PyBDSF\footnote{https://www.astron.nl/citt/pybdsf/} Python package (here VLASS and our dedicated observations) we are already provided intra-catalog connections, which can be drawn immediately. This is especially powerful given that these particular catalogs represent the deepest observations of many of these fields, giving us extra insight into sources with more complex morphology. From here, we look for inter-catalog connections using a likelihood ratio metric, where sources are connected if their uncertainty normalized distance

\begin{equation}
    d_R = \sqrt{ \left( \frac{\Delta\alpha}{\sigma_\alpha} \cos{\bar{\delta}}\right)^2 + \left( \frac{\Delta\delta}{\sigma_\delta} \right)^2} \lessapprox 3.71 \text{,}
\end{equation}

where $\Delta\alpha$ and $\Delta\delta$ represent the difference in right ascension and declination respectively, and the uncertainties here are the quadrature sums $\sigma_i^2 = \sigma_{i,1}^2 + \sigma_{i,2}^2$. Note that the fractional uncertainty on right ascension is scaled by the cosine of the mean declination to remove coordinate effects. This criterion originally comes from \citet{1977A&AS...28..211D}, and is derived by assuming the sources are scattered as a Rayleigh distribution. The choice of cutoff at $\sqrt{2\ \ln{10^3}}\approx 3.71$, inspired by a similar cutoff applied in the creation of the LOTSS DR1 catalog \citep{2017A&A...598A.104S}, is set such that we would miss only 1 in 1000 sources randomly scattered in such a distribution, while also not making the search radius so large that we have a large chance of encountering a separate source. 

\begin{figure*}
    \centering
    \includegraphics[width=\linewidth]{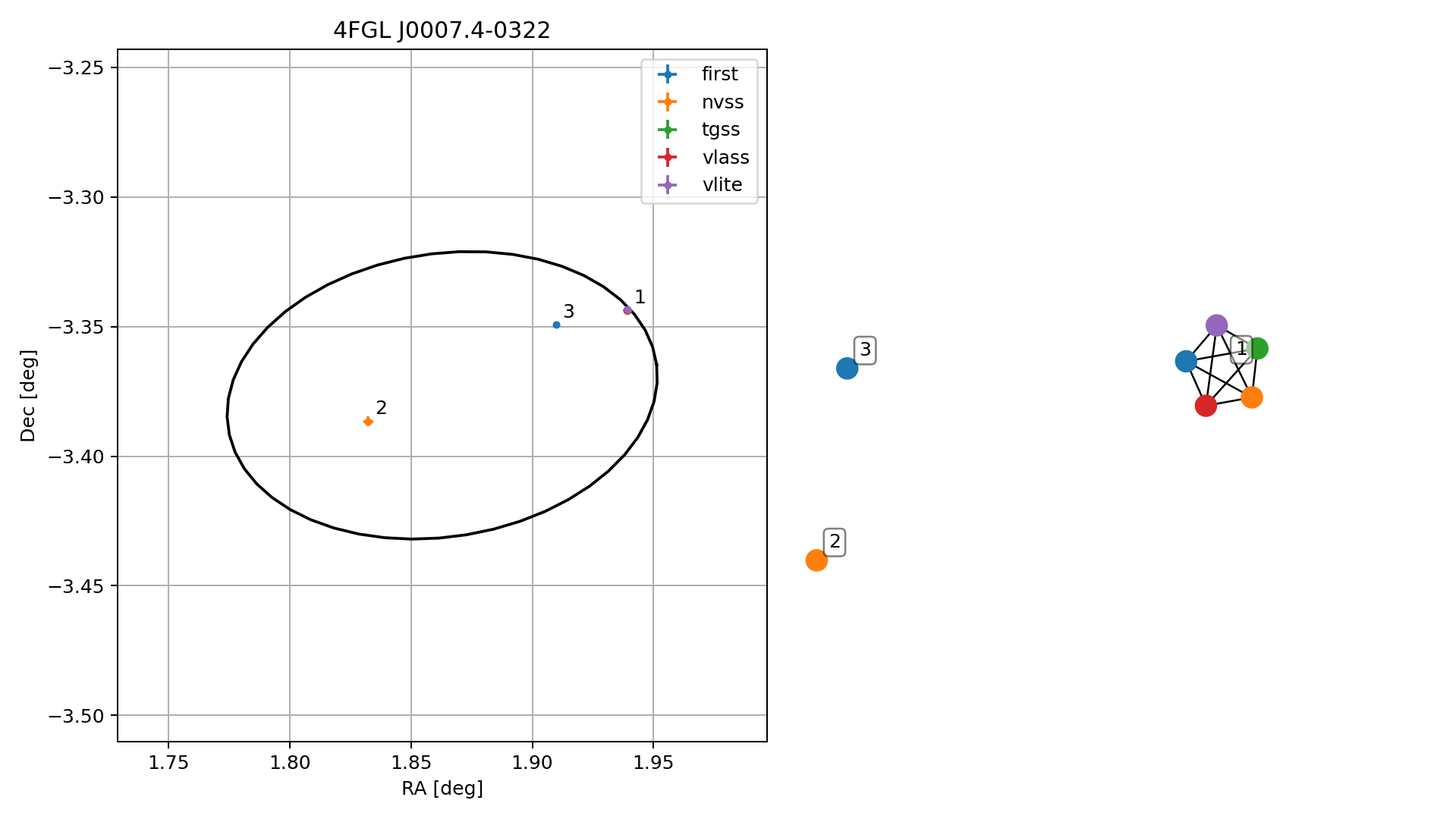}
    \caption{An unassociated field with a few well defined sources, one of which can easily be identified in multiple catalogs. On the left we show the coordinates of each radio source, along with the properties of the \textit{Fermi} ellipse. On the right is a more abstract view of the source graph used to illustrate the catalog composition of each source. We provide these plots for every \textit{Fermi} source whose positional uncertainty ellipse contains at least one radio source, along with small guide table which one can use to find the sources in the larger Multi-Survey Catalog or their respective original radio catalogs. Note that for Source 1, there are in fact a number of sources shown in the left-hand image, they are simply plotted over-top of each other. }
    \label{fig: field}
\end{figure*}

Treating each source inside the \textit{Fermi} ellipse as a node in a graph and each connection as an edge between two nodes, we can then extract connected component sub-graphs using the networkx python package. An example field, featuring two simple sources and one more complicated networked source, is shown in Figure \ref{fig: field}. Every one of these sub-graphs represents a multi-frequency source which we can extract the spectrum of. The general function we fit to our data is of the form

\begin{equation} \label{eqn: flux}
    S(\nu) = S_0 e^{\left( \alpha_0 x + c_0 x^2 \right)} \text{,}
\end{equation}

where $x=\ln{\left({\nu}/{\nu_0}\right)}$. In this case we are fitting for the reference flux $S_0$, the spectral index $\alpha_0$, and the spectral curvature $c_0$. We can choose the reference frequency $\nu_0$ which we perform the fit at, and so we adopt the logarithmic midpoint of the source data such that $\nu_0 = \sqrt{\nu_{min} \cdot \nu_{max}}$, as this seems to minimize the final uncertainty. Of course not every source will contain enough unique frequency points to be fit by this function, so we apply the following procedure:

\begin{itemize}
    \item 2 Frequencies - calculate values and uncertainties directly from the data points.
    \item 3 Frequencies - Fit a version of Equation \ref{eqn: flux} where $c_0=0$. 
    \item 4+ Frequencies - Fit the full version of Equation \ref{eqn: flux} to the data.
\end{itemize}

\begin{figure}
    \centering
    \includegraphics[width=\linewidth]{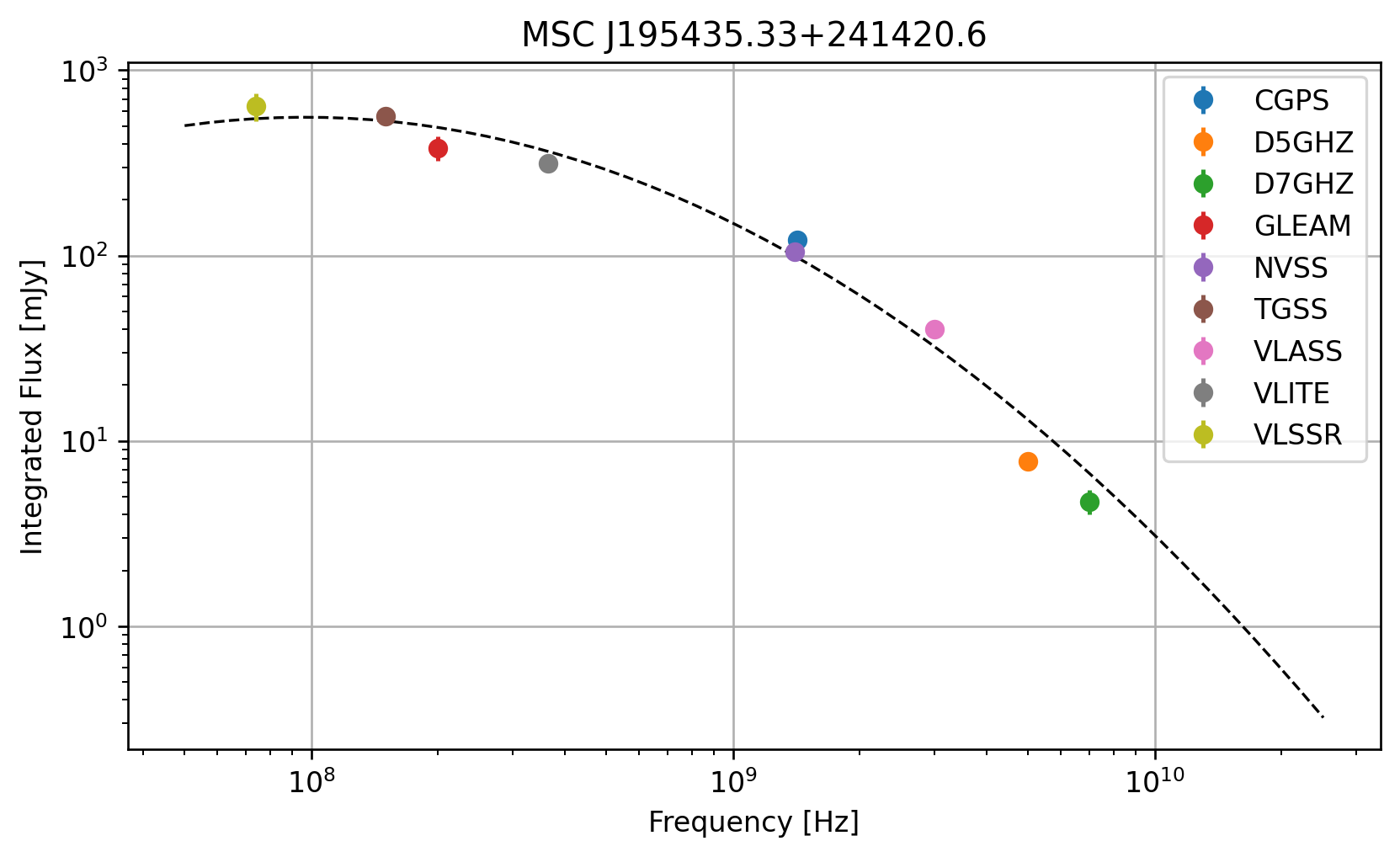}
    \caption{A sample spectrum which covers nearly our entire range of frequencies, in good agreement with the fit. The \textit{Fermi} source containing this radio source is unassociated.}
    \label{fig: spectra}
\end{figure}

These fits are performed using the \texttt{curve\_fit}\footnote{\url{https://docs.scipy.org/doc/scipy/reference/generated/scipy.optimize.curve_fit.html}} method in Scipy, which also provides the parameter uncertainties. These fit parameters and their uncertainties are then recorded along with the catalog entry names from their respective catalogs. Figure \ref{fig: spectra} shows an example of a well constrained fit of a source with fairly obvious spectral curvature, which appears in a variety of catalogs.

Where multiple sources from the same catalog are present (e.g. sources linked by PyBDSF in VLASS), we calculate the sum of their fluxes, and provide the name of the source with the lowest positional uncertainty. We note that this should be a sufficient solution for point-like or minimally resolved sources, which are the primary targets of this study. Finally we generate a weighted average of the sky coordinates, and from that produce a unique IAU-style name for each source.

\begin{deluxetable*}{lccl}
	\tablecaption{MSC Column Description}
	\label{tbl: msccols}
	\tablehead{ \colhead{Name} & \colhead{Format} & \colhead{Unit} & \colhead{Description}} 
	\startdata
	Name            &       & 23s   & Generated positional name \\
	FermiName       &       & 17s   & Fermi field which contains this source \\
	FermiClass      &       & 05s   & Fermi object classification \\
	RA              & deg   & .4f   & Weighted average of Right Ascension (J2000) \\
	RAErr           & deg   & .2e   & Weighted error on average RA \\
	Dec             & deg   & .5f   & Weighted average of Declination (J2000) \\
	DecErr          & deg   & .2e   & Weighted error on average Dec \\
	Name\_AT20G\_05 &       & 20s   &  \\
	Name\_AT20G\_08 &       & 20s   &  \\
	Name\_AT20G\_20 &       & 20s   &  \\
	Name\_CGPS      &       & 19s   &  \\
	Name\_D5GHZ     &       & 05s   &  \\
	Name\_D7GHZ     &       & 05s   &  \\
	Name\_FIRST     &       & 16s   &  \\
	Name\_GLEAM     &       & 14s   &  \\
	Name\_LOTSS     &       & 22s   &  \\
	Name\_NVSS      &       & 14s   &  \\
	Name\_PMN       &       & 15s   &  \\
	Name\_SUMSS     &       & 20s   &  \\
	Name\_TGSS      &       & 24s   &  \\
	Name\_VLASS     &       & 23s   &  \\
	Name\_VLITE     &       & 07s   &  \\
	Name\_VLSSR     &       & 22s   &  \\
	Name\_WENSS     &       & 16s   &  \\
	Name\_WISH      &       & 18s   &  \\
	RefFreq         & Hz    & .4e   & Reference frequency at the log-midpoint of data \\
	RefFlux         & mJy   & .3f   & Fit flux value at RefFreq \\
	RefFluxErr      & mJy   & .3e   & Fit uncertianty on RefFlux \\
	Alpha           &       & .3f   & Fit spectral index, positive convention \\
	AlphaErr        &       & .3e   & Fit uncertianty on Alpha \\
	Curve           &       & .3f   & Fit spectral curvature \\
	CurveErr        &       & .3e   & Fit uncertianty on Curve \\
        Notes           &       &  13s  & Notes on sources \\
	\enddata
	\tablecomments{Here the format column refers specifically to the Python string formatting code used to generate the final output table.}
\end{deluxetable*}

This data, along with various metadata on the ellipse containing each source, is then combined into our final data product, which we call the Multi-Survey Catalog (MSC). This catalog is provided alongside this article and can also be found (along with diagnostic plots such as that shown in Figure \ref{fig: field}) at \url{www.cv.nrao.edu/F357/MSC/}. Table \ref{tbl: msccols} describes the structure of the catalog. The intent of this catalog is to provide a high level overview of radio sources in a given \textit{Fermi} field, which can then be used to select for targets of interest based on properties such as flux density, spectral index, or curvature in a given band. Transformation of the given values to a new reference frequency is straightforward, with flux being scaled as in Equation \ref{eqn: flux}, and spectral index scaling like

\begin{equation}
    \alpha(\nu) = \alpha_0 + 2 c_0 x \text{.}
\end{equation}

Note that because spectral curvature is the highest-order spectral term we fit for, the curvature term will be the same regardless of the choice of frequency. If further details are needed, our catalog then identifies the relevant sources in the various radio catalogs which may be of interest. This catalog may then be used as the basis for further analysis of these unassociated fields, some of which we detail in the section to follow.

\section{Analysis} \label{sec: analysis}

As a test of our methods, we performed the above processing on all sources in the 4FGL-DR3 catalog, including those which have an existing association or identification. In total we identify 37041 distinct radio sources inside of the positional error ellipses of 5746 \textit{Fermi} fields (see Section \ref{sec: discuss} for a discussion why this number is lower than the 6659 total objects in the DR3 catalog). Looking at just the unassociated fields we find 12746 radio sources inside 1622 (out of 2157) fields. Of these 3598 are identified at more than one frequency, allowing to fit for spectral index, and 1329 have enough data points to fit for spectral curvature. 

In the sections to follow, we attempt to illustrate some of the ways in which this catalog may be used to select for interesting targets. We provide some degree of analysis and discussion on each method, and where possible we also provide the relevant subset catalogs alongside the primary catalog. We by no means intend for this list of methods to be definitive, and would encourage others to find novel ways to extrapolate targets from the information provided.

\subsection{Pulsar Candidates} \label{subsec: psr-candidates}

The most obvious way we can begin to find interesting sources from our data is to look at the radio properties of the identified sources. In particular the most obvious choice is the use of the spectral index, which in this study is particularly powerful given the extended range of our spectral coverage compared to our prior efforts, especially at lower frequencies. From studies such as \citet{1971BAAS....3R.447C} we know that the bulk of extra-galactic radio sources (such as AGN) have spectral indices of $\alpha\approx-0.5$. Furthermore we know that pulsars most often have spectral indices well outside the norm, typically in the range $\alpha<-1$, with the distribution peaked at $\alpha=-1.4$ \citep{2013MNRAS.431.1352B}. 

Beginning with the pulsars, it is fairly straightforward to select for targets with the correct spectral index. As our fits include curvature, one must calculate the spectral index at a particular choice of frequency; here we choose 1.4 GHz, the frequency of the VLA L-band, as that is a likely band to use in the case of follow-up for pulsar candidates. We then select for sources with an L-band spectral index in the range $-3<\alpha<-1$, as that roughly matches the distribution shown in \citet{2013MNRAS.431.1352B} and similar works, while also excluding the majority of flatter spectrum sources and any outliers which have artificially steeper indices. This selection alone produces a total of 1213 radio sources. 

These sources are contained in 591 unassociated \textit{Fermi} fields, implying approximately 2 pulsar candidates per field, but in actuality the difference between these numbers is largely driven by a small number of fields containing a larger number (in one case as many as 40) candidates. We note that such fields do not appear to be correlated to the size of the \textit{Fermi} ellipse, but do seem to most often appear in or near the galactic plane, which could easily cause an increase in noise and the generation of spurious sources. A simple solution is then to only target sources in fields having less than some number of candidates, assuming fields with more than this number can't be trusted. As an example, we find 907 steep spectrum sources in 566 unassociated fields each containing at most 5 of these radio candidates. 

One would typically then take the further step of selecting for sources which would be bright enough to be easily observable at the chosen frequency. Interestingly, for the choice of 1.4 GHz, all radio sources in unassociated fields within this range of spectral index have spectral flux densities about 1 mJy, our typical threshold for such a cut. One such radio source can be seen in Figure \ref{fig: spectra}, showing a wide range of spectral data and easily bright enough to be considered a good choice for follow-up. We highlight these as sources which may have been passed over by traditional timing surveys due to highly scattered pulse profiles.

A further step we can take is to make use of \gammaray properties which have been known to be indicative of pulsars. One primary example comes from the 4FGL-DR3 paper, where it is noted that pulsars seem to have significant spectral curvature in the \gammaray band, and can be disentangled from blazars by their location in the significance-curvature space \citep[see Figure 15 of][]{2022ApJS..260...53A}. With this in mind we select for unassociated sources having an average significance (column Signif\_Avg in the catalog) greater than 10, and falling above the line 

\begin{equation}
    \text{LP\_SigCurv} > 0.6 \left( \text{Signif\_Avg} \right)^{0.7} \text{.}
\end{equation}

\begin{figure*}
    \centering
    \includegraphics[width=\linewidth]{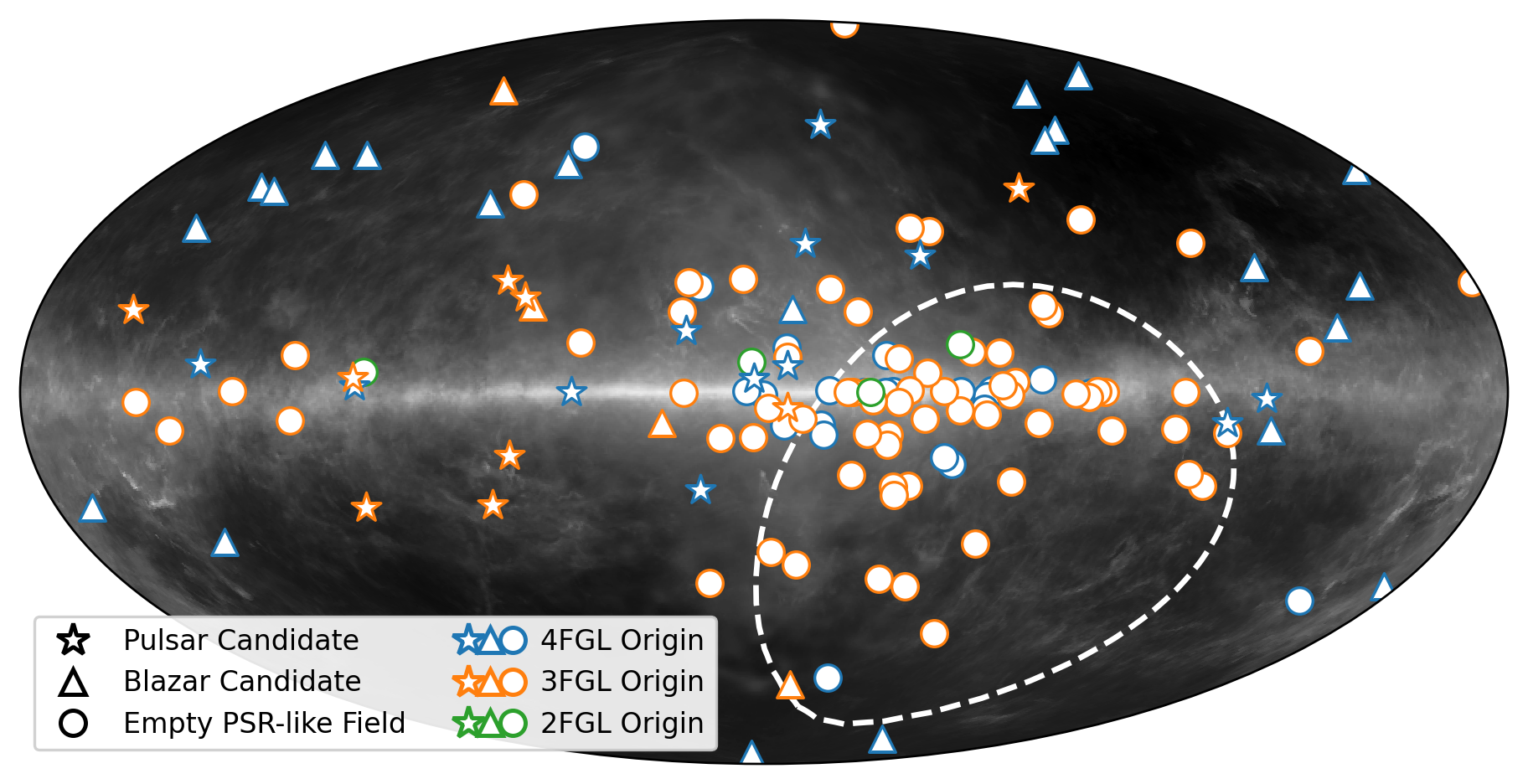}
    \caption{A galactic skymap of sources of interest discussed in Section \ref{sec: analysis}. We have overlaid these sources on the Fermi 4FGL-DR3 sensitivity map for the sake of illustration. The dashed-white line represents the minimum observable Declination of the VLA, meaning sources inside the circle will generally have been out of reach of Northern observatories. The plot features 21 Pulsar Candidates, 26 Blazar Candidates, and 97 Empty PSR-like Fields.}
    \label{fig: prettysources}
\end{figure*}

This line was selected by eye to divide the associated and identified populations of blazars and pulsars. The cutoff in significance keeps our selection out of the region of the parameter space where the distinction is more ambiguous. This particular selection is of course hardly novel on its own, but gains more leverage when collated with the radio selected pulsar-like sources. Here we enforce the further criteria that we're only interested in sources which are moderately well defined spectrally, such that their fractional spectral index error is less than 50\%. With these \gammaray and radio criteria in hand, we identify 21 sources which we designate as pulsar candidates of high interest, making ideal targets for follow-up. These 21 sources are noted in the MSC and Table \ref{tbl: interesting} with the "psr-candidate" note, as well as illustrated in Figure \ref{fig: prettysources}.

\subsection{Blazar Candidates}

We also note that the above methodology can be extended to flat spectrum sources if one is searching instead for AGN. By selecting for bright and flat spectrum radio sources, primarily outside the galactic plane, one might be able to identify a suitable list of potential AGN candidates worthy of follow-up via VLBI or a similar method. As flat-spectrum sources make up the bulk of radio sources, one would likely require further criteria to generate a reasonably small list of targets. 

For example, one could easily make us of a machine learning generated catalog such as those in \citet{2016MNRAS.462.3180C} or \citet{2016ApJ...820....8S}. Here we compare our data against the 134 blazar candidates identified by \citet{2019ApJ...887...18K}, which used \gammaray and likely X-ray properties as inputs for both a decision tree and random forest machine-learning classification system. We identify 26 of these blazar-like candidate fields as containing a flat spectrum radio source, all of which would thus be prime candidates for immediate follow-up, and notably we're able to do this using only existing catalogs, without the need to further survey these fields. All 26 are illustrated in  Figure \ref{fig: prettysources}. 20 of these sources are noted in the MSC and Table \ref{tbl: interesting} with the "blz-candidate" note. 

We note the remaining 6 sources separately, as they also appear in \citet{2022arXiv220810015K}. This more recent work performed further multi-regime classification in an attempt to identify blazar candidates as one of two primary subtypes, either flat spectrum radio quasars (FSRQs) or BL-Lac type objects. All 6 of the flat spectrum radio sources we found are located inside fields classified as BL-Lac candidates, and as such are denoted as "bll-candidates" in the MSC and in Table \ref{tbl: interesting}. 

This method could easily be extended to other 

\startlongtable
\begin{deluxetable*}{llrrrc}
\tablecaption{Targets of Interest}
\tablehead{\colhead{Name} & \colhead{FermiName} & \colhead{Spectral Index} & \colhead{LP SigCurv} & \colhead{Signif Avg} & \colhead{Candidate Type}}
\startdata
MSC J015905.20+331257.6 & 4FGL J0159.0+3313 & -0.55 & 0.36 & 8.84 & bll \\
MSC J040921.74+254442.0 & 4FGL J0409.2+2542 & -0.86 & 1.21 & 7.39 & bll \\
MSC J080056.56+073235.0 & 4FGL J0800.9+0733 & -0.66 & 3.56 & 9.12 & bll \\
MSC J083855.81+401736.1 & 4FGL J0838.5+4013 & -0.84 & 0.43 & 4.13 & bll \\
MSC J091429.68+684508.4 & 4FGL J0914.5+6845 & -0.68 & 0.37 & 8.56 & bll \\
MSC J155734.66+382030.1 & 4FGL J1557.2+3822 & -0.51 & 1.50 & 4.00 & bll \\
MSC J003655.87-265215.2 & 4FGL J0037.2-2653 & -0.71 & 0.41 & 4.20 & blz \\
MSC J013724.50-324039.3 & 4FGL J0137.3-3239 & -0.16 & 1.16 & 5.18 & blz \\
MSC J040607.94+063922.0 & 4FGL J0406.2+0639 & -0.93 & 0.12 & 4.22 & blz \\
MSC J072310.80-304754.0 & 4FGL J0723.1-3048 & -0.44 & 0.34 & 8.24 & blz \\
MSC J073725.92+653637.8 & 4FGL J0737.4+6535 & -0.83 & 1.23 & 6.66 & blz \\
MSC J075615.95-051254.5 & 4FGL J0755.9-0515 & -0.67 & 1.76 & 8.26 & blz \\
MSC J090605.54-100907.6 & 4FGL J0906.1-1011 & -0.93 & 0.58 & 6.84 & blz \\
MSC J093420.67+723045.1 & 4FGL J0934.5+7223 & -0.96 & 0.76 & 7.43 & blz \\
MSC J104705.92+673758.0 & 4FGL J1047.2+6740 & -0.39 & 1.36 & 7.85 & blz \\
MSC J104938.78+274213.0 & 4FGL J1049.8+2741 & -0.67 & 0.05 & 6.51 & blz \\
MSC J111147.61+013907.2 & 4FGL J1111.4+0137 & -0.70 & 0.00 & 4.44 & blz \\
MSC J111444.14+122618.1 & 4FGL J1114.6+1225 & -0.61 & 3.08 & 4.48 & blz \\
MSC J112213.70-022913.8 & 4FGL J1122.0-0231 & -0.20 & 1.93 & 7.75 & blz \\
MSC J122438.56+701649.7 & 4FGL J1224.6+7011 & -0.72 & 0.97 & 8.28 & blz \\
MSC J125636.31+532549.4 & 4FGL J1256.8+5329 & -0.51 & 0.79 & 5.64 & blz \\
MSC J162331.34-231334.3 & 4FGL J1623.7-2315 & -0.78 & 1.08 & 4.68 & blz \\
MSC J164825.28+483656.4 & 4FGL J1648.7+4834 & -0.98 & 1.65 & 5.71 & blz \\
MSC J181805.97+253548.0 & 4FGL J1818.5+2533 & -0.60 & 2.80 & 6.64 & blz \\
MSC J185559.92-122329.2 & 4FGL J1856.1-1222 & +0.75 & 1.37 & 9.30 & blz \\
MSC J232718.97-413437.5 & 4FGL J2326.9-4130 & -0.68 & 0.67 & 6.60 & blz \\
MSC J020509.38+665340.5 & 4FGL J0204.7+6656 & -1.78 & 4.67 & 10.97 & psr \\
MSC J053314.10+594509.3 & 4FGL J0533.6+5945 & -1.56 & 8.50 & 16.77 & psr \\
MSC J075149.66-293021.7 & 4FGL J0752.0-2931 & -1.17 & 6.55 & 11.18 & psr \\
MSC J075158.77-293602.9 & 4FGL J0752.0-2931 & -1.22 & 6.55 & 11.18 & psr \\
MSC J075451.90-395313.3 & 4FGL J0754.9-3953 & -1.43 & 6.80 & 12.37 & psr \\
MSC J120325.67-175001.9 & 4FGL J1203.5-1748 & -1.05 & 3.31 & 10.71 & psr \\
MSC J120342.65-174918.7 & 4FGL J1203.5-1748 & -1.24 & 3.31 & 10.71 & psr \\
MSC J135638.90+023722.6 & 4FGL J1356.6+0234 & -1.41 & 4.73 & 10.52 & psr \\
MSC J140715.01-301456.6 & 4FGL J1407.7-3017 & -1.51 & 5.35 & 10.17 & psr \\
MSC J140725.95-301445.7 & 4FGL J1407.7-3017 & -1.54 & 5.35 & 10.17 & psr \\
MSC J152953.52-151835.0 & 4FGL J1530.0-1522 & -1.34 & 5.18 & 10.01 & psr \\
MSC J171109.30-300536.7 & 4FGL J1711.0-3002 & -1.39 & 6.72 & 13.21 & psr \\
MSC J173526.34-071803.9 & 4FGL J1735.3-0717 & -1.33 & 4.14 & 13.51 & psr \\
MSC J173928.23-253112.0 & 4FGL J1739.3-2531 & -1.23 & 4.16 & 11.95 & psr \\
MSC J174652.52-350528.7 & 4FGL J1747.0-3505 & -1.41 & 3.88 & 10.14 & psr \\
MSC J180550.22+340116.8 & 4FGL J1805.7+3401 & -1.81 & 4.38 & 14.30 & psr \\
MSC J181341.68+282008.7 & 4FGL J1813.5+2819 & -1.20 & 4.60 & 11.88 & psr \\
MSC J190835.87+081523.4 & 4FGL J1908.7+0812 & -1.57 & 7.56 & 13.74 & psr \\
MSC J194019.07-251552.0 & 4FGL J1940.2-2511 & -1.13 & 3.90 & 13.59 & psr \\
MSC J202631.53+143054.0 & 4FGL J2026.3+1431 & -1.30 & 4.52 & 11.35 & psr \\
MSC J210744.58+515736.6 & 4FGL J2108.0+5155 & -2.20 & 4.81 & 14.29 & psr \\
MSC J210803.98+515253.8 & 4FGL J2108.0+5155 & -1.30 & 4.81 & 14.29 & psr \\
MSC J211437.06+502155.0 & 4FGL J2114.3+5023 & -1.02 & 3.74 & 12.60 & psr \\
MSC J211654.13+134155.8 & 4FGL J2117.0+1344 & -1.65 & 4.31 & 12.74 & psr \\
MSC J211709.60+134416.4 & 4FGL J2117.0+1344 & -1.22 & 4.31 & 12.74 & psr \\
MSC J225020.87+330429.7 & 4FGL J2250.5+3305 & -1.20 & 7.51 & 17.10 & psr
\enddata
\end{deluxetable*}
 \label{tbl: interesting}

\subsection{Super-Empty Fields}

The concept of Empty Fields in this context was first discussed in \citet{2017ApJ...838..139S}, which noted that a number of \textit{Fermi} fields contained no significant ($S_\nu>2\ \text{mJy}$) radio sources between 4 and 10 GHz. These fields represent an interesting subset of unassociated sources where instead of any sort of ambiguity in potential associations, we simply have no likely sources which could be producing the \gammarays. 

For the MSC it would be more difficult to establish a specific criterion for significance in terms of spectral flux density, as our frequency range is much wider and we're particularly interested in sources with steeper spectra. Instead we choose to highlight what we dub Super-Empty Fields: unassociated \textit{Fermi} fields which contain no radio sources whatsoever. We identify 537 such fields, a catalog of which (MSC\_empty.fits) is provided among our data products. 

There are several explanations for these sorts of sources, all of which make them somewhat tantalizing targets for follow-up. To begin with, a large fraction likely exist owing to a lack of coverage in the Southern Hemisphere, as can be illustrated by examining the large number residing at low Declination and unreachable by Northern observatories. These fields provide motivation toward surveys of unassociated sources by Southern observatories (e.g. ATCA, MeerKAT, SKA, ASKAP), which should shed new light on a significant number of these fields.

Of course not all of these fields are due to gaps in coverage, as shown by the numerous fields comfortably reachable in the Northern sky. This implies the existence of a population of \gammaray sources across the sky which remain resistant against traditional search methods. We can explain these sources two ways: either the sources do not produce radio emission, or the emission is in some way obscured or complicated.

Toward the first explanation, there has been interest in these fields as possible sites of dark matter annihilation. It is believed that WIMP-like dark matter would self-annihilate along channels which would produce \gammarays in a range observable to \textit{Fermi} and/or \textit{HESS} \citep[see][respectively]{2019JCAP...07..020C,2018PhRvL.120t1101A}. As the annihilation signal is confined to high energies, one would not expect any sort of complimentary signal at lower energies, especially in the radio. As these sources would be by their nature effectively invisible to radio telescopes, the main insight we can provide is identifying and associating radio/\gammaray sources toward completeness of the catalog, leaving only dark-matter candidates to study. 

If however these fields simply have had their radio emission missed by searches thus far, then there are a few potential origins. For fields in the galactic plane, it is possible that these could represent pulsars which have had their pulsed emission scattered by intervening medium, effectively wiping out the pulsed signal which would typically identify them in pulsation searches. These scattered pulsars would be observable by other methods, such as their steep spectrum, and so can be targets by dedicated low-frequency observations, or by archival methods such as those illustrated in this paper. It is also known that a majority of pulsars exhibit high circular polarization in their continuum emission \citep{1991MNRAS.253..377K}, and thus this could be used as further evidence toward a pulsar target. Once such targets are found, dedicated timing observations can be performed and used to definitively identify the \gammaray source. 

We note that all of the above methods have found some degree of success in follow-up observations. One particular example to highlight is the recent discovery of PSR J0002+6216 \citep{2019ApJ...876L..17S}, a cannonball pulsar having been found in one of the empty fields listed in \citet{2017ApJ...838..139S}, which has not beforehand been identified by pulsation searches. This system is set to provide significant insight into the evolution of pulsar wind nebulae (Kumar et al. 2022, in prep.), as well as the pulsar natal-kick velocity distribution (Bruzewski et al. 2022, in prep).

Pulsars can also provide an explanation for some fields outside the galactic plane. It is generally noted that \gammaray pulsars typically fall into one of two populations: either recycled MSPs, or young pulsars \citep{2013ApJS..208...17A}. For the first type, the binary in which these have been recycled will have had time to migrate out of the plane, and can be easily missed by blind pulsation searches, as binaries significantly complicate the process of searching for periodic signals. As such we compare our super-empty fields and MSC sources against \textit{Gaia} DR3 binaries identified in \citet{2022arXiv220606032G} as having a compact component. While we do not find any cross matches to MSC radio sources, we do note 22 super-empty fields containing at least one of these binaries, and highlight them in the catalog. 

Finally, these super-empty fields, especially those outside the galactic plane, may harbor high redshift radio galaxies (HzRGs). While the exact magnitude of this effect is still in question \citep[see][for such discussion]{2019ApJ...883L...2M,2021ApJ...911..120C,2021MNRAS.505.1543H}, it is thought that at higher redshifts inverse Compton scattering of accelerated particles off CMB photons may become the dominant cooling mechanism over the synchrotron emission typically seen in more local populations of AGN. This would generally lead to a dimming of radio emission seen from these objects at increasing redshifts, as well as an enhancement in the X-rays (where the CMB photons are up-scattered to). This effect has been noted as a possible explanation for the apparent lack of radio-loud AGN at high redshifts \citep{2011MNRAS.416..216V}. With this is mind, a \gammaray source outside of the galactic plane could be explained by such an HzRG, where the \gammarays have reached us unimpeded, but the radio emission one would typically expect has been quenched.

The degree of this quenching is again a matter of active discussion, but is not expected to wholly remove the radio luminosity in its entirety. Thus the best way to probe such fields towards these objects would be deep observations, which could potentially pick out the weaker-than-expected radio flux. It is of some note that the expected continuum spectrum for these objects likely falls into a similar steep-spectrum range as that expected for pulsars, and as such there are ongoing efforts to determine ways to adequately disentangle the two. As an example, the \gammaray properties of these populations might be expected to separate similarly to what was described previously for blazars/pulsars, and so one could conceivably look for Super-Empty Fields with pulsar-like \gammaray properties, meeting the criteria established in Section \ref{subsec: psr-candidates}. We identify 97 of these pulsar-like Super Empty Fields, which we then illustrate in Figure \ref{fig: prettysources}. 

\section{Discussion \& Conclusions} \label{sec: discuss}

The objective of this work has been to highlight the utility of existing data toward the study of unassociated \gammaray sources. In this process we have generalized an approach for networking catalogs into multi-frequency sources, providing immediate and extended insight into the characteristics of the sources inside these fields. This method catalogs nearly 13000 unique radio sources among the unassociated fields, a large number of which are detected in multiple catalogs. This spectral information has been used to generate various lists of interesting targets, and it is our intent that our Multi-Survey Catalog serve as a stepping-off point for others to generate their own targets of interest. In the prior sections we also particularly highlight the utility of our enhanced low frequency coverage toward picking out steep-spectrum objects, which is of particular note if one seeks to find pulsar associations for a number of the unassociated fields (or in some cases HzRGs, as discussed above). 

\begin{figure}
    \centering
    \includegraphics[width=\linewidth]{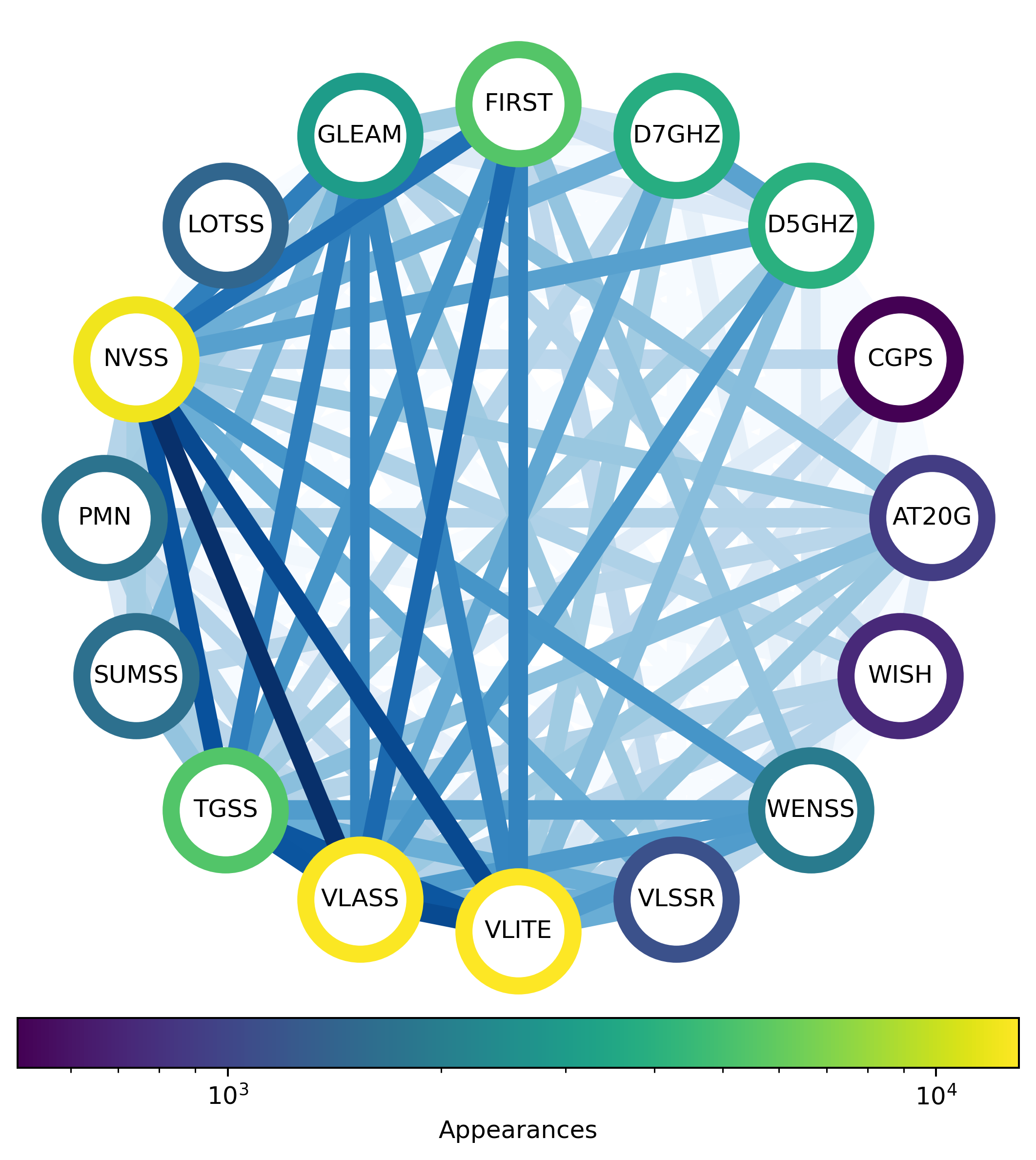}
    \caption{Connection network showing which catalogs feature most prominently and how often certain catalogs appear together. Mostly a selection of the deepest catalogs, biased toward all-sky coverage in the Northern Hemisphere. The color of each catalog node represents the number of occurrences of a source from that catalog in the MSC, while the lines between the nodes are shaded to approximate the number of of times a given pair of catalogs appear together.}
    \label{fig: connections}
\end{figure}

There is also interesting insight to be gleaned from the networking itself. Figure \ref{fig: connections} shows the inter-connectivity of the various catalogs which were used in our processing, showing which catalogs appear most frequently (which effectively probes the relative scale and depth of said surveys) as well as which catalogs appear most frequently together (probing overlaps in coverage, ideally covering different frequencies so as to provide unique information). The dominance of northern-sky catalogs, as well as their high degree of connectivity, illustrates one of the larger gaps in our understanding of these unassociated sources, namely the lack of all-sky surveys in the Southern hemisphere. We also note the relative lack of complete high frequency all-sky catalogs, with AT20G being the singular exception. For a number of the sources described in the sections prior, high frequency data points or even a confirmed non-detection at those frequencies can provide significant leverage over the spectral index of a source. Future surveys with coverage in the Southern hemisphere, such as C-BASS \citep{2018MNRAS.480.3224J} or RACS \citep{2021PASA...38...58H}, should lend significant insight into these fields. RACS especially could serve as a comparable lever-arm in the southern hemisphere, perhaps analogous to VLASS in the Northern Hemisphere, and as such further analysis incorporating this survey is planned.

Our analysis was confined to the positional uncertainty ellipses of \textit{Fermi} unassociated sources, which would rarely contain more than 100 sources across the various catalogs. Generation of the network graph is effectively an $\mathcal{O}(n^2)$ operation, as it calculates the distances between all pairs of sources. One could conceivably extend this methodology to larger regions (or even the entirety) of the sky, at the cost of greatly increased computational complexity, or via the use of more advanced pair-finding algorithms (such as some sort of uncertainty aware k-d tree). As astronomy moves towards larger data products and more complete catalogs of the radio sky, such further refinements and extensions to this methodology may prove invaluable, both to the subject of unassociated fields and beyond. 

\section{Acknowledgements}

We thank Dale Frail for useful discussions in the context of this paper, as well as Emil Polisensky for his contribution of early VLITE data toward this analysis. SB, FKS, and GBT, acknowledge support by the NASA Fermi Guest Investigator program, grants 80NSSC19K1508, NNH17ZDA001N, NNX15AU85G, NNX14AQ87G, and NNX12A075G. The National Radio Astronomy Observatory is a facility of the National Science Foundation operated under cooperative agreement by Associated Universities, Inc. Support for this work was provided by the NSF through the Grote Reber Fellowship Program administered by Associated Universities, Inc./National Radio Astronomy Observatory.

This work has made use of data from the European Space Agency (ESA) mission {\it Gaia} (\url{https://www.cosmos.esa.int/gaia}), processed by the {\it Gaia} Data Processing and Analysis Consortium (DPAC, \url{https://www.cosmos.esa.int/web/gaia/dpac/consortium}). Funding for the DPAC has been provided by national institutions, in particular the institutions participating in the {\it Gaia} Multilateral Agreement.

This research has made use of NASA’s Astrophysics Data System and has made use of the NASA/IPAC Extragalactic Database (NED) which is operated by the Jet Propulsion Laboratory, California Institute of Technology, under contract with the National Aeronautics and Space Administration. This research has made use of data, software and/or web tools obtained from NASA’s High Energy Astrophysics Science Archive Research Center (HEASARC), a service of Goddard Space Flight Center and the Smithsonian Astrophysical Observatory, of the SIMBAD database, operated at CDS, Strasbourg, France.

\software{\href{www.astropy.org}{Astropy} \citep{2022ApJ...935..167A},
          \href{www.matplotlib.org}{matplotlib} \citep{MatplotlibCitation},
          \href{www.networkx.org}{networkx} \citep{NetworkxCitation},
          \href{www.numpy.org}{Numpy} \citep{NumpyCitation},
          \href{www.scipy.org}{Scipy} \citep{ScipyCitation}}

\bibliography{bibfiles/auto,bibfiles/manual}{}
\bibliographystyle{aasjournal}

\end{document}